%% file: main.tex
\pdfoutput=1
\documentclass{article}
\usepackage[T1]{fontenc}
\usepackage{spconf,amsmath,graphicx,booktabs,multirow,url,array}
\usepackage[table]{xcolor}
\graphicspath{{figures/}}

\newcommand{\sysname}{AcoustiClaim}

\definecolor{qwenaudio}{HTML}{E4E4E4}
\definecolor{qwenomni}{HTML}{FCEFC2}
\definecolor{granitesp}{HTML}{D3E4F7}
\definecolor{audioflam}{HTML}{E6D8F3}
\definecolor{geminifl}{HTML}{D9EFDD}
\definecolor{oursband}{HTML}{FAD4D0}

\makeatletter
\long\def\@makecaption#1#2{\vskip\abovecaptionskip
 \setbox\@tempboxa\hbox{#1. #2}
 \ifdim \wd\@tempboxa >\hsize #1. #2\par \else
 \hbox to\hsize{\hfil\box\@tempboxa\hfil}\fi}
\def\thebibliography#1{\section{References}\list
 {[\arabic{enumi}]}{\settowidth\labelwidth{[#1]}\leftmargin\labelwidth
 \advance\leftmargin\labelsep
 \setlength{\itemsep}{0pt}\setlength{\parsep}{0pt}\setlength{\topsep}{2pt}%
 \usecounter{enumi}}
 \def\newblock{\hskip .11em plus .33em minus .07em}
 \sloppy\clubpenalty4000\widowpenalty4000
 \sfcode`\.=1000\relax}

\makeatother

\title{AcoustiClaim: A Numeric Claim Benchmark with Instrument Ground Truth}

\name{Sheng-Tse Lin, Siyuan Zhai$^{\ast}$, Chien-Liang Kuo$^{\ast}$, Massa Baali, Bhiksha Raj\thanks{\normalsize $^{\ast}$Siyuan Zhai and Chien-Liang Kuo contributed equally.}}
\address{Carnegie Mellon University}

\begin{document}
\ninept
\raggedbottom
\maketitle

\begin{abstract}
Audio language models state numbers for acoustic quantities, and neither human opinion nor a judge model says whether such a number is true of the signal. \sysname{} extracts each numeric claim from free text, scores it against the instrument that defines the quantity, and classes each quantity by where its reference can be read. Four open-weight systems and one closed model, asked for ten quantities five ways on two corpora, fill 207 cells. Of these, 49 emit fewer than five distinct values, and eight of the 158 cells that can be ranked exceed a rank correlation of $0.3$, the bar we set, three with an interval clear of it, five of them one closed model reading pitch. Error sits at or above a constant-predictor floor in every ranked cell but three. The reference decoder we train declines the five voice quantities in prose on $95\%$ of mixtures, with nothing withheld, and states them on the clean twins, reproducing its targets' rule from audio alone. With a calibrated threshold, withholding lowers error on all ten quantities on the mixtures in the mean and on eight at every split, against at most $0.6\%$ from a random selector. A linear baseline orders errors at least as well as ours. F0 s.d. and shimmer stay above the constant floor.
\end{abstract}


\input{sections/01_introduction}
\input{sections/02_experiments}
\input{sections/04_conclusion}

\begin{samepage}
\noindent\textbf{Compliance with ethical standards.} Every clip comes from LibriSpeech or AMI, both released for research. No participants were recruited and no new audio was recorded. The mixing manifests, the five elicitation prompts, the parser, the scorer and the generated outputs are at \url{github.com/sheng-tse/acousticlaim}, and no audio is redistributed. Some figure elements are BioRender content. Created in BioRender. Lin, S. (2026).
\par
\end{samepage}

\bibliographystyle{IEEEbib}
\bibliography{references}

\end{document}

%% file: sections/01_introduction.tex
\section{Introduction}
\label{sec:intro}

\input{figures/fig_exchange}

\looseness=-1 A speech model asked to describe a recording will often answer with numbers \cite{chu2024qwen2audio}, an SNR of 14 dB, a jitter of 1.5\%, or four pauses. Each has an established estimator, and a reader takes it as a measurement of the audio in front of the system.

\input{tables/tab_tiers}

\looseness=-1 Existing evaluations check these numbers against opinion scores or a judge model. ALLD \cite{chen2025alld} reports smaller regression models outperforming many audio language models on numerical MOS. QualiSpeech \cite{wang2025qualispeech} scores descriptions against human annotations and CoLMbo \cite{baali2025colmbo} against a speaker embedding. The benchmarks nearest ours, SALMon \cite{maimon2025salmon}, AIR-Bench \cite{yang2024airbench}, MMAU \cite{sakshi2025mmau} and MUSE \cite{carone2026muse}, score answers to labelled questions. Non-intrusive estimators return one scalar from the signal alone, an opinion score \cite{itu2004p563,reddy2021dnsmos,mittag2021nisqa} or a reference metric read without its reference \cite{kumar2023squim}, while this work scores free prose and asks whether each reference is there at all.

\looseness=-1 We score a numeric acoustic claim against the instrument that defines the quantity \cite{boersma2001praat,falk2010srmr}. In a controlled mixture we also hold the source signals the reference is read on, so Table~\ref{tab:tiers} sorts the ten quantities by whether each reference is present in the model's input.

\looseness=-1 We contribute the benchmark and its parser, which reads 13 quantities out of free text (Fig.~\ref{fig:exchange}), a reference system that carries the tiering, and three findings. (1) A cell is one system stating one quantity at one of five elicitation rungs on one corpus, on at least 25 clips. Of 207 such cells, 49 emit under five distinct values. Eight of the 158 we can rank exceed the $0.3$ bar we set. Five of the eight are the closed model reading pitch, so three of the five systems clear the bar, one if the five whose interval contains the bar are set aside. Error sits at or above the constant-predictor floor in every ranked cell but three, and each system fails in its own way. (2) With nothing withheld, the reference decoder declines the five voice quantities in prose on $95\%$ of mixtures and states them on the clean twins. Its linear head makes each stated value an exact sum of terms over the prefix tokens. Fig.~\ref{fig:mapfield} sets that decomposition beside the frames the instrument reads. Three quantities identify their own mixture above chance, HNR at $6.06$ times. (3) With a calibrated threshold we withhold $26.9\% \pm 2.8$ of slots. Error falls on all ten quantities on the mixtures in the mean and on eight at every split. A random selector at the same coverage moves error by at most $0.6\%$.

%% file: figures/fig_exchange.tex
\begin{figure}[t]
\centering
\includegraphics{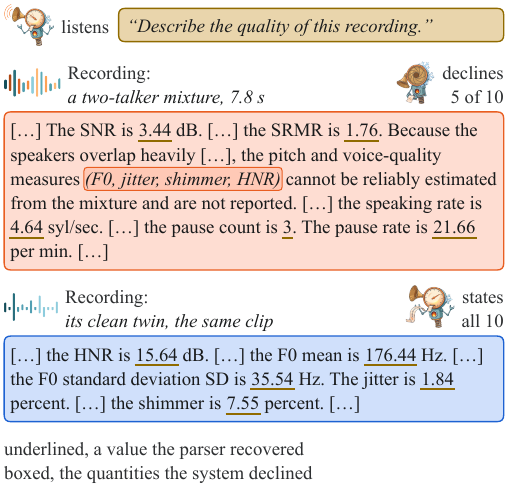}
\caption{One recording decoded twice by our system, on the two-talker mixture and on its clean twin.}
\label{fig:exchange}
\end{figure}

%% file: tables/tab_tiers.tex
\begin{table*}[t]
\centering
\setlength{\abovecaptionskip}{0pt}
\caption{Spearman correlation against the instrument, readout A on a 300-mixture, five-speaker Libri2Mix subset, beside the matched ridge. The five rows marked $\ddagger$ are read on the clips' clean twins, the mixtures set aside, and $\dagger$ is a reference read outside the overlap windows detected on the mixture, so the two twin F0 rows carry the mixture's own windows and Tier is the tier on the mixture. Case (i) sums and (ii) quotients, Eq.~\eqref{eq:algebra}, Stated the share of the scored clips, the twins on $\ddagger$ rows, the decoder speaks on, and $\pm$ a sample s.d. over five training seeds. Marks, here and in Table~\ref{tab:panel}: a bracket an nMAE at or above the floor, a dash under 25 parsed claims, c under five distinct values, excl a reference we do not score on that corpus, $\S$ a cell read on part of the clips, $\P$ two training seeds, and a red arrow a value over the bar. Praat's two half-stem readings agree at $0.42$, $0.19$ and $0.54$ on jitter, shimmer and HNR, $0.60$, $0.32$ and $0.70$ stepped up to full length, which caps any correlate at $0.77$, $0.57$ and $0.84$; jitter's $+0.809$ exceeds its cap and its within-speaker $+0.646$ does not. The other seven are a mixing parameter, deterministic or masked, so split-half does not apply.}
\label{tab:tiers}
\vspace{3pt}
\setlength{\tabcolsep}{6pt}
\renewcommand{\arraystretch}{0.95}
\begin{tabular*}{\textwidth}{@{\extracolsep{\fill}} l l c r r r r r}
\toprule
Quantity & Tier & Case & Stated & Ours $\rho$ $\pm$ s.d. & Ridge $\rho$ & Ours nMAE & Ridge nMAE \\
\midrule
SRMR & Exact & (ii) & $1.00$ & $+0.918 \pm 0.003$ & $+0.881$ & \hphantom{[}$0.394$\hphantom{]} & \hphantom{[}$0.473$\hphantom{]} \\
SNR & Measurable & (ii) & $1.00$ & $+0.972 \pm 0.005$ & $+0.944$ & \hphantom{[}$0.198$\hphantom{]} & \hphantom{[}$0.274$\hphantom{]} \\
Speaking rate & Stem-only & (i) & $1.00$ & $+0.497 \pm 0.030$ & $+0.650$ & \hphantom{[}$0.881$\hphantom{]} & \hphantom{[}$0.802$\hphantom{]} \\
Pause count & Stem-only & (i) & $1.00$ & $+0.504 \pm 0.019$ & $+0.465$ & \hphantom{[}$0.812$\hphantom{]} & [$1.059$] \\
Pause rate & Stem-only & (i) & $1.00$ & $+0.409 \pm 0.043$ & $+0.470$ & \hphantom{[}$0.888$\hphantom{]} & \hphantom{[}$0.902$\hphantom{]} \\
F0 mean$^{\ddagger}$ & Masked$^{\dagger}$ & (ii) & $0.95$ & $+0.434 \pm 0.029$ & $+0.482$ & \hphantom{[}$0.876$\hphantom{]} & \hphantom{[}$0.829$\hphantom{]} \\
F0 s.d.$^{\ddagger}$ & Masked$^{\dagger}$ & (ii) & $0.95$ & $+0.165 \pm 0.044$ & $+0.187$ & [$1.037$] & [$1.074$] \\
Jitter$^{\ddagger}$ & Stem-only & (ii) & $0.99$ & $+0.809 \pm 0.012$ & $+0.813$ & \hphantom{[}$0.709$\hphantom{]} & \hphantom{[}$0.713$\hphantom{]} \\
Shimmer$^{\ddagger}$ & Stem-only & (ii) & $0.99$ & $+0.481 \pm 0.033$ & $+0.528$ & [$1.082$] & \hphantom{[}$0.927$\hphantom{]} \\
HNR$^{\ddagger}$ & Stem-only & (ii) & $0.99$ & $+0.795 \pm 0.012$ & $+0.811$ & \hphantom{[}$0.582$\hphantom{]} & \hphantom{[}$0.585$\hphantom{]} \\
\bottomrule
\end{tabular*}
\end{table*}

%% file: sections/02_experiments.tex
\section{Methodology}
\label{sec:method}

\subsection{Benchmark}
\label{sec:benchmark}

\looseness=-1 We score on two corpora. Libri2Mix two-talker mixtures \cite{cosentino2020librimix,panayotov2015librispeech} that we reverberate with recorded room responses \cite{ko2017reverberant} and mix with noise at an independent SNR, and AMI distant-microphone meetings \cite{carletta2005ami}. Half the clip or more is overlapped in $99.1\%$ of the $3{,}000$ test mixtures and in $2.7\%$ of AMI clips. The Libri2Mix SNR reference is the level of the injected noise against the speech, and the two talkers' ratio is a separate quantity we do not score. Each mixture contributes two clips, itself and its clean twin, so the test split is $6{,}000$. Our system trains on Libri2Mix train-clean-100 alone, so its AMI rows are zero-shot, as the panel's are. Every AMI reference is read on the distant-microphone channel that the models hear, and no level was injected there. The AMI SNR reference is a within-clip dynamic range, the 90th over the 10th percentile of frame energy. Every system heard raw waveforms truncated to 10\,s, unnormalised.

\looseness=-1 In Table~\ref{tab:tiers}, Exact means the reference is fixed by the signal the model hears, Measurable a mixing parameter, Stem-only read on the clean stem the model never hears, and Masked read on the mixture outside the detected overlap windows. Silero VAD on the two separated stems fixes those windows, keeping intersections of at least $0.1$\,s. The two F0 references are Masked and marked $\dagger$, so the interferer leaks in wherever the detector misses, and the mask is strict, leaving 76 of 606 voiced frames on the example stem. On AMI the reference is read on that same channel, so its voice and timing quantities are Exact and carry the room. Exact says what fixes the reading, not that a distant-channel voice reference is phonation alone.

\looseness=-1 The parser reads 13 quantities out of generated prose (Fig.~\ref{fig:exchange}) by quantity name and unit; a description that states no value lowers coverage. Three of the 13 are not instrument readings, so we score ten.

\looseness=-1 Let $\hat{y}_i$ and $y_i$ be the claimed and the measured value on clip $i$ of $n$, and $S$ a kept subset. We report
\setlength{\abovedisplayskip}{4pt}\setlength{\belowdisplayskip}{4pt}
\begin{equation*}
\begin{aligned}
c(S) &= |S|/n, \qquad R(S) = |S|^{-1}\textstyle\sum_{i \in S} \bigl|\hat{y}_i - y_i\bigr|,\\
\mathrm{nMAE}(S) &= \frac{R(S)}{|S|^{-1}\sum_{i \in S} \bigl|\bar{y} - y_i\bigr|},
\end{aligned}
\end{equation*}
and the Spearman correlation $\rho$ over $S$. Praat reads the two F0 rows over a $75$ to $500$\,Hz pitch range at a $10$\,ms step. Jitter and shimmer come off a point process over the same range, local and in percent, and HNR off a $75$\,Hz floor. The three timing rows come off an intensity contour at a $50$\,Hz minimum pitch and a $-25$\,dB silence threshold. Here $c$ is the coverage, $R$ the selective risk \cite{geifman2017selective} and $\bar{y}$ a constant predictor. In both tables that constant is the median of the reference over the clips scored and is recomputed on $S$ wherever a threshold applies, so a cell emitting one value cannot score below $1.00$. A within-speaker correlation centres both sides on the speaker's own mean.

\looseness=-1 Three thresholds govern the external panel, $n \geq 25$ clips, at least five distinct values and $\rho > 0.3$. Under the first a cell reads as a dash in Table~\ref{tab:panel} and under the second as c, with no rank correlation to report. We set the bar ourselves and read our own rows against it. Three AMI columns read excl, the two pause references degenerate there and the SNR reference a different quantity.

\subsection{Reference system}
\label{sec:system}

\input{tables/tab_panel}

\looseness=-1 The reference system is a frozen WavLM encoder \cite{chen2022wavlm} read at layer 7, a convolutional compressor and an adapter, and a Qwen3-8B decoder under a LoRA \cite{hu2022lora} of rank 16, one prefix token per 160\,ms. One linear head over the $P$ pooled prefix tokens $\mathbf{z}_t$, the confidence head, emits a value $v_q$ and a log-variance $s_q$ per quantity. We train it by the heteroscedastic Gaussian negative log-likelihood \cite{kendall2017uncertainties} in units of that quantity's dispersion, one constant per quantity fixed before training from the training split's dispersion, with the value gradient detached and the error term re-weighted.

\looseness=-1 Each value the head states is an exact sum of per-token terms $a_{q,t}$. Three quantities are sums or rates over the $T$ frames at a known scale $\alpha$ with per-frame error $\varepsilon_k$, and seven are quotients $U/V$ of two estimated aggregates with errors $\delta U$ and $\delta V$, where $\hat{V}$ is the estimate of $V$. The two cases fix the clip error,
\begin{equation}
\begin{aligned}
v_q &= \textstyle\sum_{t=1}^{P} a_{q,t} + b_q, \quad a_{q,t} = \mathbf{w}_q^{\top}\mathbf{z}_t / P,\\
\text{(i)}\;\; \hat{y} - y &= \alpha \textstyle\sum_{k=1}^{T} \varepsilon_k, \quad
\text{(ii)}\;\; \hat{y} - y = (V \delta U - U \delta V)/(V \hat{V}),
\end{aligned}
\label{eq:algebra}
\end{equation}
so a sum accumulates evidence and its clip error, flat in the clip length for a rate and linear for a count, is at most $|\alpha| T \max_k |\varepsilon_k|$, while a quotient's is unbounded.

\looseness=-1 A claim is stated when its predicted log-variance is at or below a per-quantity threshold, $S_q = \{\, i : s_{q,i} \leq \tau_q \,\}$, and withheld otherwise. We calibrate each $\tau_q$ on a held-out half to a $0.75$ coverage target and apply it to the evaluation half, over five calibration splits on one training seed, whose spread we report as a population s.d. One quantity on one clip is a slot, ten scored per clip, and the withholding rate those thresholds realise is the cut. We hold the per-frame overlap channel at zero, so the run hears audio alone. The training descriptions decline those five once the labelled overlap passes $0.5$, so the $95\%$ is that rule recovered from audio rather than read off an input.

\looseness=-1 Readout A is the emitted text at the decoder's own coverage, on 600 and 839 Libri2Mix clips and 859 AMI clips; readout B is the confidence head, at the calibrated cut on each split's evaluation half in Fig.~\ref{fig:pair}(a) and over the full test set in Fig.~\ref{fig:pair}(b). Readout A's 600 clips are 300 mixtures and their clean twins and fill Table~\ref{tab:tiers}, which reads the mixtures alone except on the five rows marked $\ddagger$, read on those clips' twins. Its 839 Libri2Mix panel clips are mixtures scored alone and its 859 AMI panel clips are meeting recordings, and the two fill Table~\ref{tab:panel}. Readout B is the confidence head, which Fig.~\ref{fig:pair} reads on the evaluation half of each calibration split, pause count on the decoder's integer.

\section{Experimental Setup}
\label{sec:setup}

\input{figures/fig_mapfield}
\input{figures/fig_pair}

\looseness=-1 The panel is four open-weight systems, Qwen2-Audio-7B-Instruct \cite{chu2024qwen2audio}, Qwen2.5-Omni-7B \cite{xu2025qwen25omni}, Granite-Speech-3.3-8B \cite{saon2025granite} and Audio-Flamingo-3 \cite{goel2025af3}, and one closed model, Gemini 3.8 Flash (gemini-3.8-flash) through its API at temperature 0 and its lowest thinking setting. We ask each for all ten quantities five ways, a free description, a request for numbers, the quantities named with units, a template and a worked example. Every open-weight system decodes greedily with 160 new tokens, as ours does, under the same references, parser and thresholds. Each of eight sampling seeds draws its own 120-clip sample under the same greedy decode, $4{,}800$ generations per system and corpus. We pool the samples by clip, so $n$ counts distinct clips carrying a parsed claim, at most the 839 Libri2Mix or 859 AMI panel clips. Truncation at the cap runs $0.7$ to $15\%$ per system, too little to explain the $55.6\%$ of generations that parse to nothing. Granite-Speech-8B reaches 25 parsed claims in no AMI cell, so its AMI row enters no count.

\section{Results and Discussion}
\label{sec:results}

\noindent\textbf{Tiers (Table~\ref{tab:tiers}).} \looseness=-1 An estimator suite on the mixture reproduces the SRMR reference exactly and recomputes jitter and shimmer with intervals containing zero. On the mixtures the two channel rows, SRMR and SNR, hold and the three timing rows sit $0.11$ to $0.18$ under their pooled values. The five voice rows are read on the twins. The matched ridge is the benchmark's one baseline, a linear probe on the same frozen encoder on the clips each row retained. It ranks better than us on seven of the ten rows, five of them by under $0.05$. In nMAE we err less on a different seven, by $0.075$ or more on the channel rows and on pause count, where the ridge sits above the floor. Within speaker the matched ridge is ahead by $0.151$ on speaking rate and $0.073$ on pause rate and within $0.04$ on the other three mixture rows.

\smallskip\noindent\textbf{Where the reference sits (Fig.~\ref{fig:mapfield}).} \looseness=-1 Each stated value is an exact sum of per-token terms, but the prefix tokens pass a context block first, so the sum is not resolved to frames. On the drawn mixture half of a reference's magnitude falls inside $0.1$ to $4$\,s, while our own decomposition holds $0.76$ to $1.30$ times an even split on each column. A diagnostic scores how often that decomposition picks its own clip out of 20 candidates, on seven of the ten quantities. SRMR has no time-resolved reference, and the two pause references are dropped before scoring, a frame-index ramp already reaching $97\%$ of the best agreement there, above the $70\%$ we allow. Three clear chance, HNR at $6.06$ times, jitter at $4.45$ and speaking rate at $4.05$. F0 s.d. reaches $1.63$, and SNR, F0 mean and shimmer sit near chance.

\smallskip\noindent\textbf{External panel (Table~\ref{tab:panel}).} \looseness=-1 On 240 generations stratified over the four open-weight systems, both corpora and three of the five rungs at one sampling seed, the parser's precision is $0.997$ and its recall $0.763$. That audit leaves out the closed model supplying five of the eight cells over the bar, and neither denominator separates a parser miss from an abstention. A lossy parser hides positives and cannot manufacture them, so the count of cells over the bar is a lower bound while the ranking of systems is not protected. Libri2Mix yields 129 cells at 25 or more parsed claims, 91 ranked and 38 under the distinct-value minimum, and AMI 78, 67 and 11. Eight rung-level cells clear the bar, 8 of the 158 that carry a rank correlation and 8 of all 207. Three are open-weight cells resting on 10, 13 and six distinct values, each at nMAE above the floor, and five are Gemini 3.8 Flash on F0, four of them on AMI over 136 to 257 distinct values. Only its three AMI F0 mean cells fall below the floor, at $0.78$ to $0.81$, and F0 mean is the quantity whose pooled correlation is mostly speaker register on our twin rows, $+0.434$ falls to $+0.093$ within speaker, a control we run on the twin reference and not on the panel's masked cells. Fisher-$z$ $95\%$ intervals contain the bar on the three open-weight cells and on two of the five closed-model cells. Only the AMI F0 mean cells clear the bar cleanly, the largest $+0.629$ at $[0.57, 0.68]$, so three systems clear the bar with those five counted, one without. Multiplicity does not produce the eight. A null cell clears the bar $7\%$ of the time at the $n = 25$ minimum, but at the ranked cells' own $n$ a null panel would put $0.3$ cells over the bar in expectation.

\smallskip\noindent\textbf{Failure modes.} \looseness=-1 Each system fails in its own way. Granite-Speech-8B invents content, at the worked-example rung returns the example's own numbers, and on AMI states a numeral in 11 of the $2{,}880$ generations its three quantity-naming rungs produce. Its own chat template carried the audio in the user turn, so the empty row is the model's behaviour; on AMI it returns the template prompt unfilled. Gemini 3.8 Flash never falls back on a constant. The only quantity it tracks is F0, at $+0.629$ on AMI and $+0.311$ on the masked Libri2Mix reference, staying at or under $+0.14$ on the other five AMI quantities. Our own AMI row falls under the bar on three of the seven scored quantities and reads $-0.438$ on speaking rate, where the reference is a meeting rate our training never sees.

\smallskip\noindent\textbf{Withholding (Fig.~\ref{fig:pair}).} \looseness=-1 The head withholds $26.4\% \pm 2.5$ of the ten scored slots pooled over mixtures and twins, and $26.9\% \pm 2.8$ on the $3{,}000$ mixtures alone with the thresholds recalibrated there, every $\pm$ a spread over five halvings of one evaluation set, and the pooled thresholds withhold $36.8\%$ on the mixtures against $16.1\%$ on the twins. Pooled, all ten quantities reduce risk on all five splits; on the mixtures all ten fall in the mean and eight on every split. A random selector at the same coverage moves risk by at most $0.6\%$, so the gain over a random selector is in the ordering. As a selector alone, a second ridge fitted out of fold to that ridge's residual orders errors at least as well as our confidence head, winning 33 of 50 quantity-and-seed comparisons on the mixtures against none for us, and 15 of 50 against our 1 pooled. As assembled systems ours attains a better risk-coverage curve than a strong linear baseline that abstains for itself on 8 of the ten quantities with the five seeds pooled at 40 speaker clusters, the advantage in the mean predictor. Fig.~\ref{fig:pair}b draws per-seed marks at two lower cluster counts, unanimous for us on three of the ten and for the ridge on one. The head withholds more where the reference is absent. On readout B eight of the ten sit under the floor recomputed on the kept clips, F0 s.d. at nMAE $1.093$ and shimmer at $1.055$ above it.

%% file: tables/tab_panel.tex
\begin{table*}[t]
\centering
\setlength{\abovecaptionskip}{0pt}
\providecommand{\abovebar}[1]{\textcolor{red}{$\uparrow$}\,\textbf{#1}}
\providecommand{\syslogo}[1]{\raisebox{-1.7pt}{\includegraphics[height=8pt]{logos/#1}}}
\caption{Largest Spearman correlation over the rungs against the instrument. Every row is parsed emitted text, the panel pooled over eight samples by clip, mixtures only, our system trained on train-clean-100 and scored here on test clips it never saw. Marks are Table~\ref{tab:tiers}'s, and the last column counts cells over the bar at every rung, six arrows here for eight such cells. Our row carries three bases, $\S$ five voice cells over four seeds on the $6\%$ of panel clips it speaks on, five unmarked at full coverage over five seeds, and $\P$ the AMI row over two; shimmer's $+0.412$ spans $[0.145, 0.623]$ on Fisher-$z$.}
\label{tab:panel}
\setlength{\tabcolsep}{1.5pt}
\renewcommand{\arraystretch}{0.88}
\newcolumntype{Q}{>{\raggedleft\arraybackslash}p{\dimexpr(\textwidth-86pt-30pt-24\tabcolsep)/10\relax}}
\begin{tabular}{@{}>{\raggedleft\arraybackslash}p{11pt}@{\hspace{3pt}}p{72pt} Q Q Q Q Q Q Q Q Q Q >{\raggedleft\arraybackslash}p{30pt}@{}}
\toprule
\multicolumn{2}{@{}l}{System} & SRMR & SNR & Speaking & Pause & Pause & F0 & F0 & Jitter & Shimmer & HNR & $>0.3$ \\
\multicolumn{2}{@{}l}{} & & & rate & count & rate & mean & s.d. & & & & / cells \\
\midrule
\multicolumn{13}{@{}l@{}}{\emph{Libri2Mix, where the two F0 columns are masked references read on the mixture, these clips having no twin}} \\
\rowcolor{qwenaudio}
\syslogo{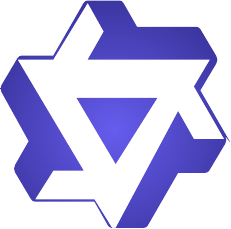} & Qwen2-Audio-7B & +0.191 & +0.024 & \textminus0.035 & \textminus0.045 & +0.050 & +0.094\hphantom{$^{\S}$} & +0.024\hphantom{$^{\S}$} & +0.012\hphantom{$^{\S}$} & +0.096\hphantom{$^{\S}$} & +0.052\hphantom{$^{\S}$} & 0/23 \\
\rowcolor{qwenomni}
\syslogo{qwen} & Qwen2.5-Omni-7B & \textminus0.057 & +0.087 & +0.021 & c & \textminus0.086 & +0.062\hphantom{$^{\S}$} & +0.009\hphantom{$^{\S}$} & +0.004\hphantom{$^{\S}$} & +0.065\hphantom{$^{\S}$} & +0.037\hphantom{$^{\S}$} & 0/29 \\
\rowcolor{granitesp}
\syslogo{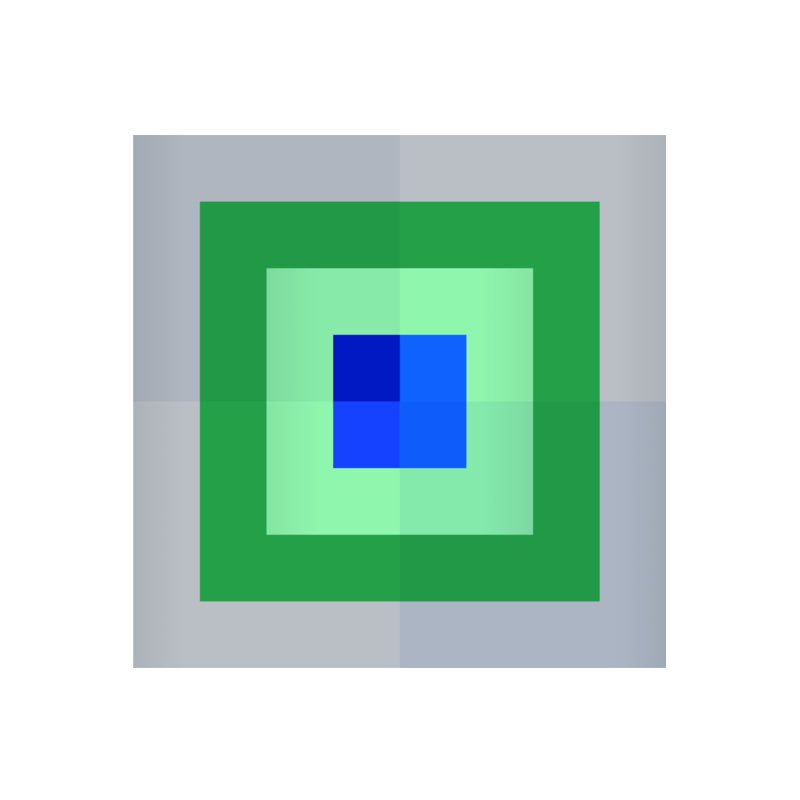} & Granite-Speech-8B & +0.047 & c & \textminus0.128 & \textminus0.097 & \textminus0.050 & c\hphantom{$^{\S}$} & c\hphantom{$^{\S}$} & \textminus0.104\hphantom{$^{\S}$} & \textminus0.063\hphantom{$^{\S}$} & +0.031\hphantom{$^{\S}$} & 0/17 \\
\rowcolor{audioflam}
\syslogo{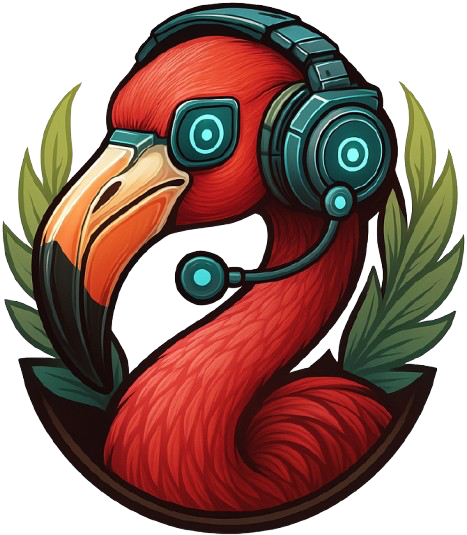} & Audio-Flamingo-3 & +0.124 & \abovebar{+0.304} & \textminus0.049 & +0.047 & +0.005 & +0.153\hphantom{$^{\S}$} & \textminus0.041\hphantom{$^{\S}$} & +0.115\hphantom{$^{\S}$} & +0.117\hphantom{$^{\S}$} & +0.018\hphantom{$^{\S}$} & \textbf{1/29} \\
\rowcolor{geminifl}
\syslogo{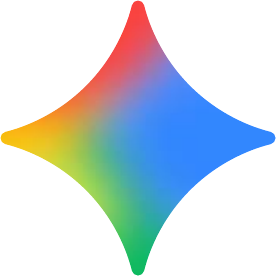} & Gemini-3.8-Flash & +0.135 & +0.229 & +0.126 & +0.205 & +0.076 & \abovebar{+0.311}\hphantom{$^{\S}$} & +0.054\hphantom{$^{\S}$} & +0.068\hphantom{$^{\S}$} & +0.059\hphantom{$^{\S}$} & +0.005\hphantom{$^{\S}$} & \textbf{1/31} \\
\addlinespace[1pt]
\rowcolor{oursband}
\syslogo{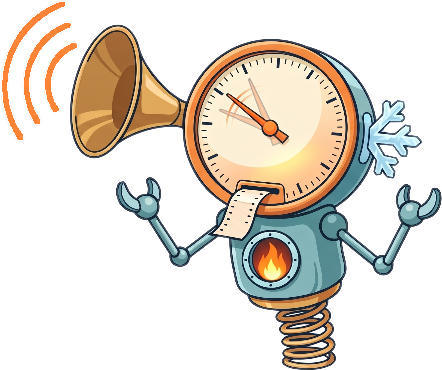} & \sysname{} & +0.942 & +0.970 & +0.360 & +0.506 & +0.336 & +0.493$^{\S}$ & +0.121$^{\S}$ & +0.507$^{\S}$ & +0.412$^{\S}$ & +0.524$^{\S}$ & ref. \\
\addlinespace[2pt]
\multicolumn{13}{@{}l@{}}{\emph{AMI, where the two pause references are degenerate and the SNR reference is a within-clip dynamic range, all three excluded}} \\
\rowcolor{qwenaudio}
\syslogo{qwen} & Qwen2-Audio-7B & +0.043 & excl & \textminus0.027 & excl & excl & +0.196\hphantom{$^{\S}$} & +0.058\hphantom{$^{\S}$} & +0.100\hphantom{$^{\S}$} & +0.096\hphantom{$^{\S}$} & \textminus0.030\hphantom{$^{\S}$} & 0/16 \\
\rowcolor{qwenomni}
\syslogo{qwen} & Qwen2.5-Omni-7B & +0.053 & excl & +0.057 & excl & excl & \abovebar{+0.381}\hphantom{$^{\S}$} & \textminus0.030\hphantom{$^{\S}$} & +0.091\hphantom{$^{\S}$} & \textminus0.014\hphantom{$^{\S}$} & +0.161\hphantom{$^{\S}$} & \textbf{1/21} \\
\rowcolor{granitesp}
\syslogo{granite} & Granite-Speech-8B & -- & excl & -- & excl & excl & --\hphantom{$^{\S}$} & --\hphantom{$^{\S}$} & --\hphantom{$^{\S}$} & --\hphantom{$^{\S}$} & --\hphantom{$^{\S}$} & 0/0 \\
\rowcolor{audioflam}
\syslogo{af3} & Audio-Flamingo-3 & +0.051 & excl & +0.253 & excl & excl & +0.044\hphantom{$^{\S}$} & +0.086\hphantom{$^{\S}$} & +0.193\hphantom{$^{\S}$} & \abovebar{+0.320}\hphantom{$^{\S}$} & \textminus0.028\hphantom{$^{\S}$} & \textbf{1/20} \\
\rowcolor{geminifl}
\syslogo{gemini} & Gemini-3.8-Flash & +0.076 & excl & +0.021 & excl & excl & \abovebar{+0.629}\hphantom{$^{\S}$} & \abovebar{+0.347}\hphantom{$^{\S}$} & +0.137\hphantom{$^{\S}$} & +0.133\hphantom{$^{\S}$} & +0.130\hphantom{$^{\S}$} & \textbf{4/21} \\
\addlinespace[1pt]
\rowcolor{oursband}
\syslogo{acousticlaim} & \sysname{}$^{\P}$ & +0.777 & excl & \textminus0.438 & excl & excl & +0.621\hphantom{$^{\S}$} & +0.194\hphantom{$^{\S}$} & +0.420\hphantom{$^{\S}$} & +0.297\hphantom{$^{\S}$} & +0.638\hphantom{$^{\S}$} & ref. \\
\bottomrule
\end{tabular}
\end{table*}

%% file: figures/fig_mapfield.tex
\begin{figure*}[t]
\centering
\includegraphics{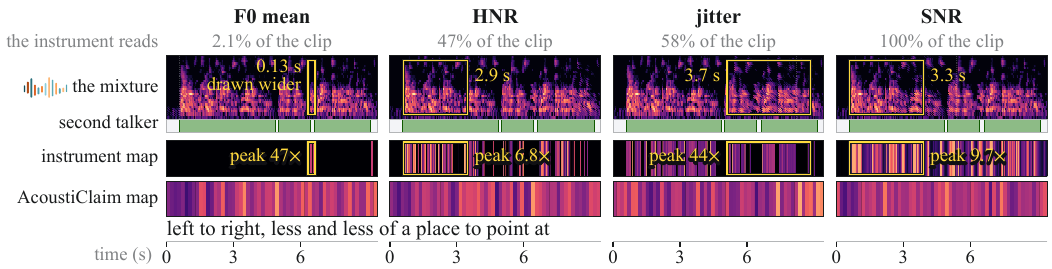}
\caption{One $9.44$\,s mixture in four columns. Rows are the mixture, the second talker's spans, the frames the instrument reads and our 58-token decomposition. Each box is the shortest window holding half the reference's magnitude and each peak a frame's height over an even split.}
\label{fig:mapfield}
\end{figure*}

%% file: figures/fig_pair.tex
\begin{figure*}[t]
\centering
\includegraphics{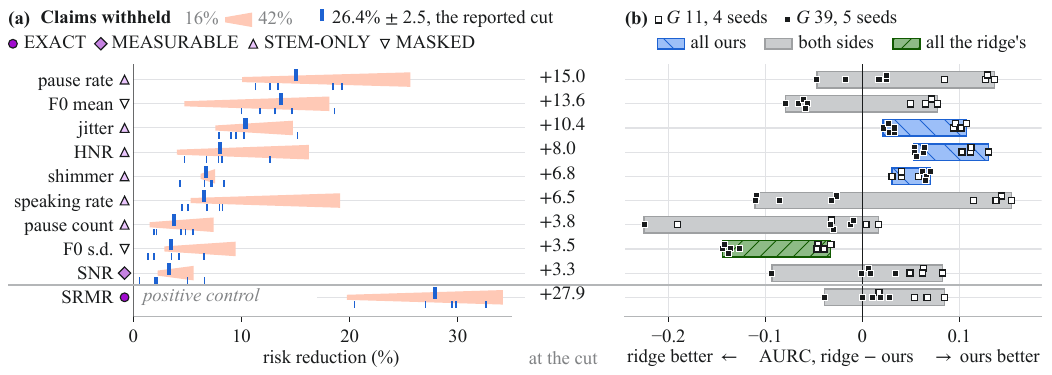}
\caption{(a) Reduction in selective risk per quantity, each wedge thickening from $16$ to $42\%$ withheld, the stroke at the reported cut, the ticks the five held-out splits, the right-hand numbers each row's reduction there, pooled over mixtures and twins, and SRMR a positive control, its reference Exact. (b) Rows as in (a), risk-coverage area (AURC), the selective risk integrated over coverage, of that residual ridge minus ours on the full test set, one mark per training seed, four at $G = 11$ and five at $G = 39$, $G$ the number of speaker clusters the area is computed over, the fifth seed scored at $G = 39$ only, each bar filled by whether the nine marks all favour one side, right of zero our win.}
\label{fig:pair}
\end{figure*}

%% file: sections/04_conclusion.tex
\section{Conclusion}
\label{sec:conclusion}

\looseness=-1 \sysname{} scores each numeric claim against the instrument that defines it and says whether that reference is in the input. Three of five systems clear the $0.3$ bar we set, in 8 of the 158 cells that can be ranked, and error sits at or above the floor in all but three. A calibrated threshold lowers error on all ten quantities on the mixtures, though the advantage over a linear baseline that abstains for itself lies in the mean predictor, not in the ordering.

%% file: main.bbl
\begin{thebibliography}{10}

\bibitem{chu2024qwen2audio}
Yunfei Chu, Jin Xu, Qian Yang, Haojie Wei, Xipin Wei, Zhifang Guo, Yichong
  Leng, Yuanjun Lv, Jinzheng He, Junyang Lin, Chang Zhou, and Jingren Zhou,
\newblock ``Qwen2-audio technical report,''
\newblock {\em arXiv preprint arXiv:2407.10759}, 2024.

\bibitem{chen2025alld}
Chen Chen, Yuchen Hu, Siyin Wang, Helin Wang, Zhehuai Chen, Chao Zhang,
  Chao-Han~Huck Yang, and Eng~Siong Chng,
\newblock ``Audio large language models can be descriptive speech quality
  evaluators,''
\newblock in {\em Proc. ICLR}, 2025.

\bibitem{wang2025qualispeech}
Siyin Wang, Wenyi Yu, Xianzhao Chen, Xiaohai Tian, Jun Zhang, Lu~Lu, Yu~Tsao,
  Junichi Yamagishi, Yuxuan Wang, and Chao Zhang,
\newblock ``{QualiSpeech}: A speech quality assessment dataset with natural
  language reasoning and descriptions,''
\newblock in {\em Proc. ACL}, 2025.

\bibitem{baali2025colmbo}
Massa Baali, Shuo Han, Syed~Abdul Hannan, Purusottam Samal, Karanveer Singh,
  Soham Deshmukh, Rita Singh, and Bhiksha Raj,
\newblock ``{CoLMbo}: Speaker language model for descriptive profiling,''
\newblock in {\em Proc. IEEE ASRU}, 2025.

\bibitem{maimon2025salmon}
Gallil Maimon, Amit Roth, and Yossi Adi,
\newblock ``{SALMon}: A suite for acoustic language model evaluation,''
\newblock in {\em Proc. IEEE ICASSP}, 2025.

\bibitem{yang2024airbench}
Qian Yang, Jin Xu, Wenrui Liu, Yunfei Chu, Ziyue Jiang, Xiaohuan Zhou, Yichong
  Leng, Yuanjun Lv, Zhou Zhao, Chang Zhou, and Jingren Zhou,
\newblock ``{AIR-Bench}: Benchmarking large audio-language models via
  generative comprehension,''
\newblock in {\em Proc. ACL}, 2024.

\bibitem{sakshi2025mmau}
S~Sakshi, Utkarsh Tyagi, Sonal Kumar, Ashish Seth, Ramaneswaran Selvakumar,
  Oriol Nieto, Ramani Duraiswami, Sreyan Ghosh, and Dinesh Manocha,
\newblock ``{MMAU}: A massive multi-task audio understanding and reasoning
  benchmark,''
\newblock in {\em Proc. ICLR}, 2025.

\bibitem{carone2026muse}
Brandon~James Carone, Iran~R. Roman, and Pablo Ripoll{\'e}s,
\newblock ``The {MUSE} benchmark: Probing music perception and auditory
  relational reasoning in audio {LLMs},''
\newblock in {\em Proc. IEEE ICASSP}, 2026.

\bibitem{itu2004p563}
{ITU-T},
\newblock ``Single-ended method for objective speech quality assessment in
  narrow-band telephony applications,''
\newblock {ITU-T} Recommendation P.563, International Telecommunication Union,
  2004.

\bibitem{reddy2021dnsmos}
Chandan K.~A. Reddy, Vishak Gopal, and Ross Cutler,
\newblock ``{DNSMOS}: A non-intrusive perceptual objective speech quality
  metric to evaluate noise suppressors,''
\newblock in {\em Proc. IEEE ICASSP}, 2021.

\bibitem{mittag2021nisqa}
Gabriel Mittag, Babak Naderi, Assmaa Chehadi, and Sebastian M{\"o}ller,
\newblock ``{NISQA}: A deep {CNN}-self-attention model for multidimensional
  speech quality prediction with crowdsourced datasets,''
\newblock in {\em Interspeech}, 2021.

\bibitem{kumar2023squim}
Anurag Kumar, Ke~Tan, Zhaoheng Ni, Pranay Manocha, Xiaohui Zhang, Ethan
  Henderson, and Buye Xu,
\newblock ``{TorchAudio-Squim}: Reference-less speech quality and
  intelligibility measures in {TorchAudio},''
\newblock in {\em Proc. IEEE ICASSP}, 2023.

\bibitem{boersma2001praat}
Paul Boersma,
\newblock ``{Praat}, a system for doing phonetics by computer,''
\newblock {\em Glot International}, vol. 5, no. 9/10, pp. 341--345, 2001.

\bibitem{falk2010srmr}
Tiago~H. Falk, Chenxi Zheng, and Wai-Yip Chan,
\newblock ``A non-intrusive quality and intelligibility measure of reverberant
  and dereverberated speech impaired by noise,''
\newblock {\em IEEE Trans. Audio, Speech, Lang. Process.}, vol. 18, no. 7, pp.
  1766--1774, 2010.

\bibitem{cosentino2020librimix}
Joris Cosentino, Manuel Pariente, Samuele Cornell, Antoine Deleforge, and
  Emmanuel Vincent,
\newblock ``{LibriMix}: An open-source dataset for generalizable speech
  separation,''
\newblock {\em arXiv preprint arXiv:2005.11262}, 2020.

\bibitem{panayotov2015librispeech}
Vassil Panayotov, Guoguo Chen, Daniel Povey, and Sanjeev Khudanpur,
\newblock ``{LibriSpeech}: An {ASR} corpus based on public domain audio
  books,''
\newblock in {\em Proc. IEEE ICASSP}, 2015.

\bibitem{ko2017reverberant}
Tom Ko, Vijayaditya Peddinti, Daniel Povey, Michael~L. Seltzer, and Sanjeev
  Khudanpur,
\newblock ``A study on data augmentation of reverberant speech for robust
  speech recognition,''
\newblock in {\em Proc. IEEE ICASSP}, 2017.

\bibitem{carletta2005ami}
Jean Carletta, Simone Ashby, Sebastien Bourban, Mike Flynn, Maël Guillemot,
  Thomas Hain, Jaroslav Kadlec, Vasilis Karaiskos, Wessel Kraaij, Melissa
  Kronenthal, Guillaume Lathoud, Mike Lincoln, Agnes Lisowska, Iain McCowan,
  Wilfried Post, Dennis Reidsma, and Pierre Wellner,
\newblock ``The {AMI} meeting corpus: A pre-announcement,''
\newblock in {\em Proc. MLMI}, 2005.

\bibitem{geifman2017selective}
Yonatan Geifman and Ran El-Yaniv,
\newblock ``Selective classification for deep neural networks,''
\newblock in {\em Proc. NeurIPS}, 2017.

\bibitem{chen2022wavlm}
Sanyuan Chen, Chengyi Wang, Zhengyang Chen, Yu~Wu, Shujie Liu, Zhuo Chen, Jinyu
  Li, Naoyuki Kanda, Takuya Yoshioka, Xiong Xiao, Jian Wu, Long Zhou, Shuo Ren,
  Yanmin Qian, Yao Qian, Michael Zeng, Xiangzhan Yu, and Furu Wei,
\newblock ``{WavLM}: Large-scale self-supervised pre-training for full stack
  speech processing,''
\newblock {\em IEEE J. Sel. Topics Signal Process.}, vol. 16, no. 6, pp.
  1505--1518, 2022.

\bibitem{hu2022lora}
Edward~J. Hu, Yelong Shen, Phillip Wallis, Zeyuan Allen-Zhu, Yuanzhi Li, Shean
  Wang, Lu~Wang, and Weizhu Chen,
\newblock ``{LoRA}: Low-rank adaptation of large language models,''
\newblock in {\em Proc. ICLR}, 2022.

\bibitem{kendall2017uncertainties}
Alex Kendall and Yarin Gal,
\newblock ``What uncertainties do we need in bayesian deep learning for
  computer vision?,''
\newblock in {\em Proc. NeurIPS}, 2017.

\bibitem{xu2025qwen25omni}
Jin Xu, Zhifang Guo, Jinzheng He, Hangrui Hu, Ting He, Shuai Bai, Keqin Chen,
  Jialin Wang, Yang Fan, Kai Dang, Bin Zhang, Xiong Wang, Yunfei Chu, and
  Junyang Lin,
\newblock ``{Qwen2.5-Omni} technical report,''
\newblock {\em arXiv preprint arXiv:2503.20215}, 2025.

\bibitem{saon2025granite}
George Saon, Avihu Dekel, Alexander Brooks, Tohru Nagano, Abraham Daniels,
  Aharon Satt, Ashish Mittal, Brian Kingsbury, David Haws, Edmilson Morais,
  Gakuto Kurata, Hagai Aronowitz, Ibrahim Ibrahim, Jeff Kuo, Kate Soule, Luis
  Lastras, Masayuki Suzuki, Ron Hoory, Samuel Thomas, Sashi Novitasari, Takashi
  Fukuda, Vishal Sunder, Xiaodong Cui, and Zvi Kons,
\newblock ``Granite-speech: Open-source speech-aware {LLMs} with strong
  {English} {ASR} capabilities,''
\newblock {\em arXiv preprint~arXiv:2505.08699}, 2025.

\bibitem{goel2025af3}
Arushi Goel, Sreyan Ghosh, Jaehyeon Kim, Sonal Kumar, Zhifeng Kong, Sang-gil
  Lee, Chao-Han~Huck Yang, Ramani Duraiswami, Dinesh Manocha, Rafael Valle, and
  Bryan Catanzaro,
\newblock ``Audio flamingo 3: Advancing audio intelligence with fully open
  large audio language models,''
\newblock in {\em Proc. NeurIPS}, 2025.

\end{thebibliography}
